\documentclass[aps,prl,twocolumn,superscriptaddress,showpacs]{revtex4}
 \usepackage{graphicx}
\newcommand{\apjl}{Astrophys. J.}      % {ApJ}
\newcommand{\aap}{Astron. Astrophys.}  % {A{\&}A}
\newcommand{\jgr}{J. Geophys. Res.}
\newcommand{\solphys}{Solar Phys.}

\begin{document}

\title{Torus instability}

\author{B. Kliem}
%\email[]{bkliem@aip.de}
\affiliation{Astrophysical Institute Potsdam, An der Sternwarte 16,
             14482 Potsdam, Germany}

\author{T. T\"{o}r\"{o}k}
\affiliation{Mullard Space Science Laboratory, Holmbury St.\ Mary,
             Dorking, Surrey RH5 6NT, UK}

\received{28 April 2006; published 26 June 2006}

\begin{abstract}

The expansion instability of a toroidal current ring in low-beta
magnetized plasma is investigated.
Qualitative agreement is obtained with experiments on spheromak expansion
and with essential properties of solar coronal mass ejections, 
unifying the two apparently disparate classes of fast and slow coronal
mass ejections.

\end{abstract}

% 52.35.Py  Macroinstabilities (hydromagnetic, e.g., kink, fire-hose, ...
% 52.30.-q  Plasma dynamics and flow
% 96.60.ph  Coronal mass ejection
% 96.60.qf  Prominence eruptions

\maketitle

\vspace*{-\baselineskip}

We consider the expansion instability of a toroidal current ring with the
goal to describe the rapid expansion of such rings or partial rings
observed in laboratory and astrophysical plasmas
\cite{Yee&Bellan2000,Forbes2000}. The equilibrium of this configuration
was established by Shafranov and is realized in the tokamak fusion device
\cite{Shafranov1966}. It necessarily includes an external poloidal
magnetic field $\mathbf{B}_\mathrm{ex}$, since the Lorentz-self force,
also referred to as the hoop force, as well as the net pressure gradient
force of a bent current channel always point radially outward.

The stability of the Shafranov equilibrium has been considered by Bateman
\cite{Bateman1978}, who found that the ring is unstable against expansion
if the external poloidal field decreases sufficiently rapidly in the
direction of the major torus radius $R$. Since the hoop force decreases
if the ring expands, a perturbation $dR>0$ will be unstable if the
opposing Lorentz force due to $\mathbf{B}_\mathrm{ex}$ decreases faster
with increasing $R$ than the hoop force. Bateman derived
% %------------------------------------------------------------------------
% \begin{equation}                                                  % Eq. .
$n=-R\,d\ln{B_\mathrm{ex}}/dR>3/2$
% \label{eq:TI_Bateman_condition}
% \end{equation}
% %------------------------------------------------------------------------
as condition for the instability, which we will refer to as the torus
instability (TI).
The TI can be regarded as a lateral kink instability distributed
uniformly over the ring. Different from the helical kink instability, the
TI cannot be stabilized by the presence of a toroidal field component
inside the torus (which occurs, for example, in a force-free
equilibrium), since the hoop force points outward also in this case, with
an $R$ dependence similar to the purely poloidal configuration.

The TI is suppressed in fusion devices by employing external poloidal
fields with sufficiently small decay indices $n$ and by stabilizing image
currents in the walls of the device. However, it may occur in
astrophysical plasmas, where the external poloidal field is often
strongly inhomogeneous \cite{Titov&Demoulin1999}, and in some plasma
experiments \cite{Yee&Bellan2000, Hansen&Bellan2001, Hsu&Bellan2003}. In
particular, the observations of erupting prominences on the Sun, which
often evolve into the cores of coronal mass ejections (CMEs) causing
major perturbations of the space weather \cite{Crooker&al.1997}, suggest
the topology of a single expanding partial current ring, whose footpoints
are anchored in the inertial visible solar surface. A threshold of $n>2$
was estimated for this case \cite{Titov&Demoulin1999}; otherwise the
instability has apparently never been reconsidered. Research on CMEs was
instead directed at the possibility of a catastrophe due to the nonexistence
of equilibrium in part of parameter space
\cite{Forbes2000,Priest&Forbes2002}.

In the present Letter, we derive a TI threshold that is somewhat more
general than the one by Batemean and treat the evolution
of the instability for the first time. We consider two cases: a freely
expanding ring relevant in the laboratory and for CMEs  and an expanding
ring with constant total current, which captures an important effect of
the footpoint anchoring on an expanding partial ring
and can be relevant in the initial stage of CMEs. We focus on the
essence of the instability and its development by including only the hoop
force (in the large aspect ratio approximation, $R\gg b$) and the
stabilizing Lorentz force due to $\mathbf{B}_\mathrm{ex}$. Gravity,
pressure, external toroidal fields, and any variation in the direction of
the minor radius $b$ are neglected to permit a largely analytical
description. The neglect of pressure effects is justified by the fact
that the instability is primarily relevant for low-beta plasmas, in which
the conversion of the stored magnetic energy is able to drive a
large-scale expansion.

% (their inclusion reduces the growth rate of the
% instability but does not alter its qualitative properties).

With these assumptions, the force balance is purely in the direction of
the major radius and given by \citep{Shafranov1966, Bateman1978}
%------------------------------------------------------------------------
\begin{equation}                                                  % Eq. 1
\rho_m\frac{d^2R}{dt^2}= \frac{I^2}{4\pi^2b^2R^2}(L+\mu_0R/2)
                        -\frac{IB_\mathrm{ex}(R)}{\pi b^2}\,,
\label{eq:TI_full}
\end{equation}
%------------------------------------------------------------------------
where $\rho_m$ is the mass density of the ring and $I$ is the total ring
current. The inductance of the ring is given by
$L=\mu_0R\left(\ln(8R/b)-2+l_\mathrm{i}/2\right)$. The internal
inductance per unit length of the ring $l_\mathrm{i}$ is of order unity
if the radial profile of the current density is not strongly peaked in
the center of the torus; in particular for uniform current density,
$l_\mathrm{i}=1/2$. The flux enclosed by the ring is
$\Psi=\Psi_I+\Psi_\mathrm{ex}$, with $\Psi_I=LI$. Ideal MHD requires
$\Psi=const$ during a perturbation $R\to R+dR$. We now have to make an
assumption how $\Psi_\mathrm{ex}$ evolves. Here we follow Bateman, who
ignored changes in the external field due to the perturbation and
evaluated the enclosed external flux using the prescribed external field
$B_\mathrm{ex}(R)$,
%------------------------------------------------------------------------
\begin{equation}                                                  % Eq. 2
\Psi=\Psi_I+\Psi_\mathrm{ex}=LI-2\pi\int_0^RB_\mathrm{ex}(r)rdr\,.
\label{eq:Psi}
\end{equation}
%------------------------------------------------------------------------
This consistency with the use of $B_\mathrm{ex}(R)$ in the expression for
the restoring force in Eq.~(\ref{eq:TI_full}) implies inconsistency
regarding the conservation of the enclosed flux. If the latter were to be
treated consistently, one would have to require
$\Psi_\mathrm{ex}(R)=const$ instead. Numerical simulations of the
instability, which will be reported elsewhere, support the instability
criterion derived from Eq.~(\ref{eq:Psi}). They also show that magnetic
reconnection sets in at the rear side of the expanding ring as the
instability develops and lets the ring effectively ``slide'' through the
external poloidal field \cite{T"or"ok&Kliem2005}, so that
Eq.~(\ref{eq:Psi}) represents a reasonable aproximation also for large
expansions. With both assumptions for $\Psi_\mathrm{ex}(R)$ it is easily
seen that the total ring current, $I(R)\le\Psi_{I0}/L(R)$, must decrease
as a free torus expands, since the logarithmic term in $L$ varies only
weakly with $R$ (subscripts 0 denote initial values here and henceforth).

We make the ansatz that $B_\mathrm{ex}(R)=\hat{B}R^{-n}$ in the region of
interest, $R\ge R_0$. (At $R\to0$ a finite $B_\mathrm{ex}$ is assumed,
whose particular value will drop out of the equations below. We also have
to assume $n\ne2$ in intermediate steps of the calculation but find that
the final expressions [right-hand side of Eq.~(\ref{eq:TI_free_norm}) and
following] match smoothly as $(n-2)\to\pm0$.) Using Eq.~(\ref{eq:Psi}),
the ring current is expressed through the initial values
%========================================================================
\begin{eqnarray}                                                  % Eq. 3
I(R)&\!=\!&\frac{c_0R_0I_0}{cR}
       \left\{1+\frac{c_0+1/2}{2c_0}\frac{1}{2-n}
                \left[\left(\frac{R}{R_0}\right)^{2-n}\!\!-1\right]\right\},
		                                             \nonumber\\*
    &\!\! & \hfill ~n\ne2\,,
\label{eq:I_free1}
\end{eqnarray}
%========================================================================
where $c=L/(\mu_0R)$. Inserting this in Eq.~(\ref{eq:TI_full}) and
normalizing, $\rho=R/R_0$ and $\tau=t/T$, where
%------------------------------------------------------------------------
\begin{displaymath}   % {equation}                                                  % Eq. 5
T=\left(
   \frac{c_0+1/2}{4}\frac{b_0^2}{B_\mathrm{eq}^2/\mu_0\rho_{m0}}\right)^{1/2}
 =\frac{(c_0+1/2)^{1/2}}{2}\frac{b_0}{V_\mathrm{Ai}}
% \label{eq:def_T_free}
\end{displaymath}   % {equation}
%------------------------------------------------------------------------
is essentially the ``hybrid'' Alfv\'en time of the minor radius (based on
the external equilibrium field $B_\mathrm{eq}=B_\mathrm{ex}(R_0)$ and the
initial density in the torus), we obtain the equation describing the
evolution of the major radius
%========================================================================
\begin{eqnarray}                                                  % Eq. 4
\frac{d^2\rho}{d\tau^2}
&=& 
 \frac{c_0^2}{(c_0+1/2)c}\,\rho^{-2}
 \left[1+\frac{c_0+1/2}{c_0}\frac{\rho^{2-n}-1}{2(2-n)}\right]
                                                             \nonumber\\*
&&
 \left\{ \frac{c+1/2}{c}
         \left[1+\frac{c_0+1/2}{c_0}\frac{\rho^{2-n}-1}{2(2-n)}\right]
 \right.                                                     \nonumber\\*
&&
 \left. ~ -\frac{c_0+1/2}{c_0}\rho^{2-n} \right\},~n\ne2\,.
 \label{eq:TI_free_norm}
\end{eqnarray}
%========================================================================
We now assume $c(R)=const$, which is exact if the expansion is
self-similar and can otherwise be expected to introduce relatively little
error because $c$ depends only logarithmically on $R/b(R)$. An
approximately self-similar evolution of a freely expanding ring has been
found in a laboratory experiment \cite{Yee&Bellan2000}, and also the
observations of CMEs indicate some degree of self-similarity
\cite{Bothmer&Schwenn1994}. With $c(R)=c_0$, the condition for
instability
$\left.d\left(d^2\rho/d\tau^2\right)/d\rho\right|_{\rho=1}>0$ becomes
%------------------------------------------------------------------------
\begin{equation}                                                  % Eq. 5
n>n_\mathrm{cr}=3/2-1/(4c_0)\,.
\label{eq:TI_free_threshold}
\end{equation}
%------------------------------------------------------------------------
Bateman's condition is recovered as $c_0\to\infty$, which may be regarded
as the ``very large aspect ratio limit.'' If
$\Psi_\mathrm{ex}(R)=\Psi_\mathrm{ex\,0}$ is assumed in
Eq.~(\ref{eq:Psi}), then the expansion is described by
$d^2\rho/d\tau^2
 =(c_0/c)^2(c+1/2)/(c_0+1/2)
  \rho^{-2}\left[1-\rho^{2-n}(c/c_0)(c_0+1/2)/(c+1/2)\right]$
instead of Eq.~(\ref{eq:TI_free_norm}), and [again with $c(R)=c_0$] the
threshold rises to $n>2$. We note that this assumption (with
$\Psi_\mathrm{ex\,0}=0$) and this threshold correspond to the case of a
gravitationally balanced current ring around a star or massive object,
which should, therefore, be marginally stable.

Equation~(\ref{eq:TI_free_norm}) can be integrated twice only for small
displacements, $0<\epsilon=\rho-1\ll1$, showing that the expansion starts
nearly exponentially,
%------------------------------------------------------------------------
\begin{equation}                                                  % Eq. 6
\epsilon(\tau)=
 \frac{v_0T/R_0}{(n-n_\mathrm{cr})^{1/2}}
 \sinh\left((n-n_\mathrm{cr})^{1/2}\tau\right),~\epsilon\ll1\,,
\label{eq:TI_free_e(t)}
\end{equation}
%------------------------------------------------------------------------
with the growth rate $\gamma=(n-n_\mathrm{cr})^{1/2}$. Here $v_0$ is
the initial velocity of the expansion resulting from a perturbation.
Integrating Eq.~(\ref{eq:TI_free_norm}) once shows that for $n>3/2$ a
constant asymptotic velocity is reached
%========================================================================
\begin{eqnarray}                                                  % Eq. 7
v_\infty
 &\!\!=\!\!&\left[\left(\frac{v_0T}{R_0}\right)^2\!+
          \frac{2(2n-3+\frac{1}{2c_0})(n-1+\frac{1}{4c_0})}
               {(2n-3)(n-1)}
     \right]^{1/2}                                           \nonumber\\*
 &\!\!\approx\!\!&\left[(v_0T/R_0)^2\!+2\right]^{1/2},
                  ~n>3/2\,.
\label{eq:TI_free_v_infty}
\end{eqnarray}
%========================================================================
For $n_\mathrm{cr}<n<3/2$, the acceleration does not decrease sufficiently
rapidly as $\rho\to\infty$ so that the asymptotic velocity diverges. This
discrepancy with the behavior at $n>3/2$ results from the simplifications
made; it would disappear if the restoring forces due to flux and pressure
pileup in front of the expanding ring, which dominate at large $\rho$,
would be included. The asymptotic gain of kinetic energy is
$\Delta W=M\int_{R_0}^\infty(d^2R/dt^2)dR\approx M(R_0/T)^2,~n>3/2$,
where $M=2\pi^2b_0^2R_0\rho_{m0}$ is the mass of the torus. 

For a large aspect ratio, the characteristic velocity in these expressions
is much larger than the hybrid Alfv\'en velocity of the initial
configuration, $R_0/T\approx(R_0/b_0)V_\mathrm{Ai}\gg V_\mathrm{Ai}$.
% 
% (with $l_\mathrm{i}\sim1/2$ we have $(c_0+1/2)^{1/2}/2\approx1$).
% 
Therefore, $v_0T/R_0\ll1$ even in the case that the initial perturbation
$v_0$ approaches $V_\mathrm{Ai}$, as may happen if it is due to a kink
instability \cite{T"or"ok&Kliem2005,T"or"ok&al.2004}. The dimensional
asymptotic expansion velocity $\approx2^{1/2}(R_0/b_0)V_\mathrm{Ai}$ for
$n>3/2$ scales as the Alfv\'en velocity of the initial configuration.

\begin{figure}[t]                                                % Fig. 1
\begin{center}
 \includegraphics[width=.43\textwidth]
                 {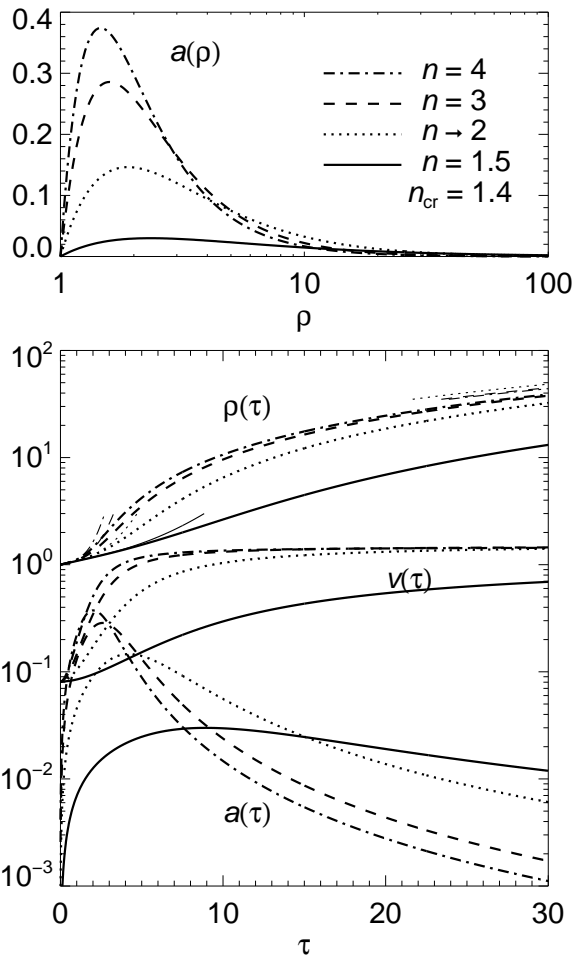}
               % {/home/kli/pap/torus1/dgl/torus_free.ps}
\end{center}
\vspace{-1.5\baselineskip}  %{-3mm}
\caption[]
{\label{fig:TI_free}
 Radial acceleration profiles and solutions of
 Eq.~(\ref{eq:TI_free_norm}) for the freely expanding torus with
 $c(R)=const$ and $R_0/b_0=10$, $v_0T/R_0=0.005$, $l_\mathrm{i}=1/2$.
 The approximate solutions, Eqs.~(\ref{eq:TI_free_e(t)}) and
 (\ref{eq:TI_free_v_infty}), are included as thin lines for $\rho(\tau)$.} 
 \vspace{-.75\baselineskip}
\end{figure}

Figure~\ref{fig:TI_free} shows the acceleration profile
$a=d^2\rho/d\tau^2$ and the numerical solution of
Eq.~(\ref{eq:TI_free_norm}) with $c(R)=c_0$, along with the analytical
approximations Eqs.~(\ref{eq:TI_free_e(t)}) and
(\ref{eq:TI_free_v_infty}), for particular values of $v_0T/R_0$ and
$R_0/b_0$ and for the practically relevant range of $n$. The acceleration
rises quickly to a maximum, which increases strongly with
$n>n_\mathrm{cr}$ and is reached within $\rho\lesssim2$ for all $n$
shown. It then decreases quickly with increasing $\rho$ for $n\gtrsim2$
but decreases only slowly for $n$ close to $n_\mathrm{cr}$. The resulting
expansion, $\rho(\tau)-1$, has an approximately exponential-to-linear
characteristic for $n\gtrsim2$ but is much closer to a
constant-acceleration curve over a considerable radial range for $n$
close to $n_\mathrm{cr}$. A qualitatively similar $n$ dependence of the
acceleration profile is obtained if
$\Psi_\mathrm{ex}(R)=\Psi_\mathrm{ex\,0}$ is assumed in
Eq.~(\ref{eq:Psi}).

%~~~~~~~~~~~~~~~~~~~~~~~~~~~~~~~~~~~~~~~~~~~~~~~~~~~~~~~~~~~~~~~~~~~~~~~~
%~~~~~~~~~~~~~~~~~~~~~~~~~~~~~~~~~~~~~~~~~~~~~~~~~~~~~~~~~~~~~~~~~~~~~~~~

This $n$ dependence of the expansion fits perfectly to the typical
characteristics of CME rise profiles. Fast CMEs reach a speed of
$\sim10^3$~km\,s$^{-1}$, comparable to the Alfv\'en velocity in the inner
corona, often within a height range of $h\lesssim R_\odot/3$ above the
photosphere and show no significant acceleration further out. These
events originate from active regions which possess a rapid decay of the
field concentration at heights comparable to the sunspot distance $D$
($D\sim R_\odot/10$ in bigger active regions); for essentially bipolar
active regions, $n>3/2$ for $h>D/2$, quickly approaching $n\approx3$ at
$h\gtrsim D$. On the other hand, slow CMEs propagate with roughly
constant, small acceleration throughout the currently observable height
range ($h\lesssim30~R_\odot$), reaching the gravitational escape speed of
a few $10^2$~km\,s$^{-1}$ typically only at heights of several $R_\odot$.
These events originate from erupting prominences far from active regions,
where the large-scale height dependence of the field, approximately
$B\propto h^{-3/2}$ \cite{Vrsnak&al.2002}, dominates already low in the
corona. Interestingly, the fastest CMEs, and the strongest flares,
originate in so-called $\delta$-spot regions, which are quadrupolar, with
one pair of opposite polarity being closely packed within a single
sunspot, so that a particularly steep field decrease ($n>3$) occurs low
in the corona within very strong fields, which imply high Alfv\'en
velocities of up to several $10^3$~km\,s$^{-1}$. Thus, the torus
instability not only provides a uniform description of the apparently
disparate classes of fast and slow CMEs \cite{MacQueen&Fisher1983} but
also explains naturally the preferred occurrence of the most powerful
solar eruptions in $\delta$-spot regions \cite{Sammis&al.2000}.

\begin{figure}[t]                                                % Fig. 2
\begin{center}
 \includegraphics[width=.43\textwidth]
                 {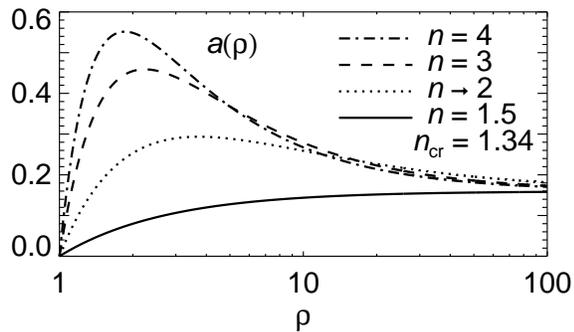}
               % {/home/kli/pap/torus1/dgl/torus_linetied_a_only.ps}
\end{center}
\vspace{-1.5\baselineskip}  %{-3mm}
\caption[]
{\label{fig:TI_tied}
 Radial acceleration profiles of the torus instability with fixed ring
 current [Eq.~(\ref{eq:TI_tied_norm})] and parameters as in
 Fig.~\ref{fig:TI_free}.}
 \vspace{-.75\baselineskip}
\end{figure}

The magnetic field in erupting prominences and CME cores can be modeled
as a section of a torus, whose remaining part is submerged and frozen in
the dense, high-beta photospheric and subphotospheric plasma. Such
line tying is generally regarded to have a stabilizing influence; for
example, in case of the helical kink instability it raises the threshold
twist from $2\pi$ to $2.49\pi$ \cite{Hood&Priest1981}. It has an even
stronger effect on the TI. If a current-carrying loop emerges or is
formed in the low corona, the line-tying is expected to suppress the
instability
completely until the loop is at least semicircular, since the major
radius of a rising loop must \emph{decrease} before that stage
\cite{Chen&Krall2003}. Beyond that point, however, the line tying
supports the expansion because it tends to keep the current through the
footpoints of the partial ring to be constant. It is
not clear at present how much of this current can enter the coronal part
of the ring, where, due to the low resistivity, reconnection cannot
easily occur within the ring so that the number of field line turns and
hence $IR$ tend to be constant. While a complete account of the
line tying requires a more sophisticated treatment, we can describe its
amplifying effect on the expansion by replacing Eq.~(\ref{eq:I_free1})
with $I(R)=I_0$, obtaining the limiting case of maximum outward
acceleration, given by
%========================================================================
\begin{eqnarray}                                                 % Eq. 8
\frac{d^2\rho}{d\tau^2}
&=& 
 \frac{1}{2(c_0+1/2)}+
 \frac{(2n-3)c_0+1/2}{2(n-2)(c_0+1/2)}\rho^{-1}              \nonumber\\*
 & & -\frac{2n-3}{2(n-2)}\rho^{1-n},~n\ne2\,.
 \label{eq:TI_tied_norm}
\end{eqnarray}
%========================================================================
The critical decay index of the external poloidal field for instability,
%------------------------------------------------------------------------
\begin{equation}                                                 % Eq. 9
n_\mathrm{cr}=3/2-1/(2c_0+1)\,,
\label{eq:TI_tied_threshold}
\end{equation}
%------------------------------------------------------------------------
is only slightly smaller than the critical index for the freely expanding
ring. The initial evolution is again given by Eq.~(\ref{eq:TI_free_e(t)}).
The strong amplifying effect becomes apparent in the further
evolution. This shows an enlarged radial range of acceleration, in better
agreement with some CME observations, and a higher peak
(Fig.~\ref{fig:TI_tied}). The asymptotic acceleration does not vanish,
however. Since $a(\rho\to\infty)$ is small only for $\ln(8R_0/b_0)\gg1$
or for $l_\mathrm{i0}\gg1$, which both do not have observational
support, it is obvious that Eq.~(\ref{eq:TI_tied_norm}) cannot hold
throughout the expansion.

%~~~~~~~~~~~~~~~~~~~~~~~~~~~~~~~~~~~~~~~~~~~~~~~~~~~~~~~~~~~~~~~~~~~~~~~~
%~~~~~~~~~~~~~~~~~~~~~~~~~~~~~~~~~~~~~~~~~~~~~~~~~~~~~~~~~~~~~~~~~~~~~~~~

\begin{figure}[t!]                                                % Fig. 3
\begin{center}
 \includegraphics[width=.43\textwidth]
                 {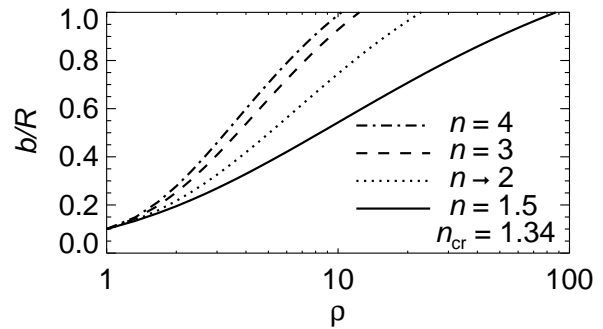}
               % {/home/kli/pap/torus1/dgl/overexpansion.ps}
\end{center}
\vspace{-1.5\baselineskip}  %{-3mm}
\caption[]
{\label{fig:overexpansion}
 Development of the inverse aspect ratio for the torus instability with
 fixed ring current [Eq.~(\ref{eq:overexpansion})] for $R_0/b_0=10$.}
 \vspace{-.75\baselineskip}
\end{figure}

Another consequence of constant ring current is the decrease of the
aspect ratio in the course of the instability. Requiring $\Psi(R)=\Psi_0$
[Eq.~(\ref{eq:Psi})] with $I(R)=I_0$, we find
%========================================================================
\begin{eqnarray}                                                 % Eq. 10
\frac{b(R)}{R}&=&\frac{8}{\exp\left\{c_0\rho^{-1}
 +\frac{c_0+1/2}{2(n-2)}\rho^{-1}(1-\rho^{2-n})+2
 -\frac{l_\mathrm{i}}{2}\right\}}                           \nonumber\\*
 & & ~n\ne2\,,
\label{eq:overexpansion}
\end{eqnarray}
%========================================================================
(Fig.~\ref{fig:overexpansion}). Such overexpansion of the minor radius is
a characteristic of many CMEs, observed as a cavity in the outer part of
the rising flux, which gives rise to the so-called three-part structure
of CMEs \cite{Crooker&al.1997}. The overexpansion is so rapid that
$b\to R$ for $\rho=R/R_0\sim10^1\mbox{--}10^2$. At this point our
simplified description breaks down. We can expect that magnetic
reconnection with the surrounding field or between the loop legs is
triggered by the overexpansion.
This implies that $I(R)=const$ no longer holds and that the
acceleration falls off as the reconnection proceeds. Comparing the
acceleration profiles in Figs.~\ref{fig:TI_free} and \ref{fig:TI_tied},
it is clear that the association of fast and slow CMEs with,
respectively, high and only slightly supercritical decay index $n$ holds
for line-tied current rings as well.

Let us finally consider the expansion of a spheromak-like torus in a
nearly field-free vacuum chamber \cite{Yee&Bellan2000}, which proceeded
in the observed range $\rho\lesssim2$ with roughly constant velocity.
We note that Taylor relaxation in the torus transformed toroidal into
poloidal flux in the course of the expansion, influencing the TI in as
yet unknown ways, and that the scatter in the data (Fig.\,19 in
Ref.~\cite{Yee&Bellan2000}) permits a fit with slightly increasing velocity
as well. With $\mathbf{B}_\mathrm{ex}=0$ and $\Psi(R)=L_0I_0$, we obtain
$d^2\rho/d\tau^2=(c+1/2)c^{-2}\rho^{-2}$ in place of
Eq.~(\ref{eq:TI_free_norm}), where time is now normalized to
$T^{\prime}=(\pi/c_0)(b_0/\tilde{V}_\mathrm{Ai})$ and
$\tilde{V}_\mathrm{Ai}$ is defined using the field ($\tilde{B}$) at $R=0$
and $\rho_{m0}$.
% 
% the field $\tilde{B}$ in the center of the ring ($R=0$) and $\rho_{m0}$.
% 
This acceleration decreases so
rapidly that, soon after onset, the expansion velocity is expected to
increase only slowly with $\rho$, consistent with the observation. The
asymptotic velocity
$\left((c+1/2)/c^2\right)^{1/2}R_0/T^{\prime}
 \sim5\mbox{--}16$~km\,s$^{-1}$,
obtained using the observed $R/b\approx2$, $\tilde{B}\sim300$~G as a
representative value of the measured range [Figs.~11, 12(b), and 12(c) in
Ref.~\cite{Yee&Bellan2000}), and estimated densities
$N\sim10^{15}\mbox{--}10^{16}$~cm$^{-3}$

%~~~~~~~~~~~~~~~~~~~~~~~~~~~~~~~~~~~~~~~~~~~~~~~~~~~~~~~~~~~~~~~~~~~~~~~~
%~~~~~~~~~~~~~~~~~~~~~~~~~~~~~~~~~~~~~~~~~~~~~~~~~~~~~~~~~~~~~~~~~~~~~~~~

\noindent
\cite{Bellan_personalcommun},
is in acceptable agreement with the observed expansion
velocity of $\approx5$~km\,s$^{-1}$.

We conclude that the TI is a possible mechanism for CMEs (in addition to
a catastrophe \cite{Forbes2000,Priest&Forbes2002} and to the helical kink
instability \cite{T"or"ok&Kliem2005}), that the TI governs their
medium-scale ($\rho\lesssim10^2$) expansion, providing a unified
description of fast and slow CMEs and a possible explanation for their
three-part structure, and that the TI occurred in experiments on
spheromak expansion.

We gratefully acknowledge constructive comments by T.G. Forbes,
P. D\'emoulin, and V.S. Titov.
This work was supported by DFG and PPARC.

% \bibliography{BKliem}

\end{document}